\documentclass[traditabstract]{aa}

\usepackage{graphicx}
\usepackage{txfonts}
\usepackage{longtable}
\usepackage{natbib}

\begin{document}

\title{Parsec-scale Properties of Brightest Cluster Galaxies.}

\author{E. Liuzzo\inst{1,2}, G. Giovannini\inst{1,2}, 
M. Giroletti\inst{1}, G.B. Taylor\inst{3,4}}
 
\institute{INAF Istituto di Radioastronomia, via Gobetti 101, 40129
Bologna, Italy \and
            Dipartimento di Astronomia, Universit\`a di Bologna, via 
Ranzani 1, 40127 Bologna, Italy \and
            Department of Physics and Astronomy, University of New 
Mexico, Albuquerque NM 87131, USA
           \and also Adjunct Astronomer at the National Radio Astronomy
Observatory, USA}

\date{Received / Accepted}

\authorrunning{E. Liuzzo, G. Giovannini, M. Giroletti, G.B. Taylor }
\titlerunning{Parsec-scale properties of BCGs.}

\abstract
{We present new VLBI observations at 5 GHz of a complete sample of
  Brightest Cluster Galaxies (BCGs) in nearby Abell Clusters (distance
  class $<$3).  Combined with data from the literature, this provides
  parsec-scale information for 34 BCGs. Our analysis of their parsec
  scale radio emission and cluster X-ray properties shows a possible
  dichotomy between BCGs in cool core clusters and those in non cool
  core clusters. Among resolved sources, those in cool core clusters
  tend to have two-sided parsec-scale jets, while those in less
  relaxed clusters have predominantly one-sided parsec-scale jets. We
  suggest that this difference could be the result of interplay
  between the jets and the surrounding medium. The one-sided structure
  in non cool core clusters could be due to Doppler boosting effects in
  relativistic, intrinsically symmetric jets; two-sided morphology in
  cool core clusters is likely related to the presence of heavy and mildly
  relativistic jets slowed down on the parsec-scale. Evidence of
  recurrent activity are also found in BCGs in cool core clusters.}

\maketitle

\keywords{galaxies: active --- galaxies: jets --- galaxies: nuclei --- galaxies: clusters: individual --- (galaxies:) cooling flows --- radio continuum: galaxies}


\section{Introduction}

Brightest Cluster Galaxies (BCGs) are a unique class of objects
\citep{ye04}. These galaxies are the most luminous and massive
galaxies in the Universe, up to ten times brighter than typical
elliptical galaxies, with large characteristic radii (tens of kiloparsecs,
\citet{scho86}). Most BCGs are cD galaxies with extended envelopes
over hundreds of kiloparsecs, but they can also be giant E and D galaxies. 
They show a very small dispersion in luminosity, making them excellent
standard candles. We refer e.g. to \citet{hoe80} for a detailed discussion
about the distribution of the absolute magnitude of BCGs which is not correlated
to the dynamic equilibrium of their host-clusters \citep{kat03}.
The
optical morphology often shows evidence of past or recent galaxy
mergers, such as multiple nuclei.  Moreover, they tend to lie very
close to peaks of the cluster X-ray emission have velocities
near the cluster rest frame velocity. All these properties
indicate that they could have an unusual formation history
compared to other E galaxies. These galaxies are intimately related to
the collapse and formation of the cluster: recent models suggest that
BCGs must form earlier, and that galaxy merging within the
cluster during collapse within a cosmological hierarchy is a viable
scenario \citep{ber06}.  BCGs in many cool core clusters often have
blue excess light indicative of recent star formation with colours
that imply starbursts occurring over the past 0.01-1 Gyr
\citep{mcn04}. Line-emitting nebulae surround approximately a third of
all BCGs \citep{cra99}, for example NGC 1275 in Perseus \citep{con01} and
A1795 \citep{cow83} exhibit extended filamentary nebulae (up to
50 kpc from the central galaxy), some of which are co-spatial with
soft X-ray filaments. Moreover, many of these BCGs also contain
reservoirs of 10$^{8}$-10$^{11.5}$M$_{\odot}$ of molecular hydrogen
\citep{ed01}.

In the radio band, BCGs have long been recognized to show peculiar
properties \citep{bu81}. They are more likely to be radio-loud than
other galaxies of the same mass \citep{bes06,dun09} and very often their
radio morphology shows evidence of a strong interaction with the
surrounding medium: some BCGs have a wide angle tail structure (WAT)
very extended on the kiloparsec-scale (e.g., 3C\,465 in A2634, \citet{sak99}),
or with small size (e.g. NGC4874 in Coma cluster, \citet{fe85}); in
other cases, we have diffuse and amorphous sources, either extended
(3C\,84 in Perseus, \citet{pe90}) or with very small size (e.g., the BCG
in A154, \citet{fe94}). These last two sources are rare in the general
radio population, but frequently present in BCGs and in particular in
BCGs located in cooling core clusters of galaxies.

Radio-loud AGN in BCGs have been proposed as a potential solution to
the cooling-flow problem \citep{mcn05}. As a consequence of radiative
cooling the temperature in the central regions of the cluster is
expected to drop. However, XMM-Newton and Chandra observations of
cooling core clusters have shown that the temperature of cluster cores
is $\sim$30$\%$ higher than expected and also the amount of cooling
gas is only about 10$\%$ of that predicted \citep{pe02}. The presence
of X-ray cavities in the emitting gas coincident with the presence of
radio lobes demonstrates the interplay between the radio activity of
BCGs and the slowing for arrest of cooling in cluster centers
\citep{bi08, dun06}. Very deep Chandra observations of the Perseus and
Virgo clusters \citep{fa03, fa06} have revealed the presence of
approximately spherical pressure waves in these clusters. These
``ripples'' are excited by the expanding radio bubbles and the
dissipation of their energy can provide a quasi-continuous heating of
the X-ray emitting gas \citep{ru04}.

Despite these results, many important properties of BCGs are poorly
known: not all BCGs are strong radio sources and cyclic activity with
a moderate duty cycle is necessary to justify the slow down of the
cooling processes in clusters where the BCG is not a high power radio
source.  Moreover, it is not clear if radio properties of BCGs in
cooling flow clusters are systematically different from those of BCGs
in merging clusters; we note that in cooling clusters the kiloparsec-scale
morphology of BCGs is often classified as a mini-halo \citep{git07,
  gov09}, but extended `normal' sources are also present (e.g., Hydra
A, \citep{tay96}), while in merging clusters the most common
morphology is a WAT source, but point-like as well as core-halo
sources are also present.

Moreover, BCGs are not yet well studied on the parsec-scale as a
class.  Only a few of them have been observed, mostly famous, bright
radio galaxies.  In some of these cases, they look like normal FRI
radio galaxies with relativistic collimated jets. Parsec-scale jets
are usually one-sided because of Doppler boosting effects (e.g.,  3C\,465
in A2634 and 0836+29 in A690, \citep{ve95}), although there are also
cases where two-sided symmetric jets are present in VLBI images, and
the presence of highly relativistic jets is uncertain (e.g. 3C\,338 in
A2199, \citep{ge07}, and Hydra A in A780, \citep{tay96}).

To investigate the properties of BCGs on the parsec-scale, we have
selected a sample of BCGs unbiased with respect to their radio and X-Ray
properties. We present the sample in \S2; in \S3 we provide
information about the observations and the data reduction; in \S4
we give notes on individual sources; in \S5 we describe the
statistical properties of our complete sample; in
\S6 we present general properties of BCGs including data from the
literature, and finally, in \S7 and \S8 are the
discussion and the conclusion respectively.

We have used the following set of cosmological parameters:
$\Omega_m$=0.3, $\Omega_\lambda$=0.7, H$_{0}$= 70 km s$^{-1}$
Mpc$^{-1}$. We define the spectral index $\alpha$ as $S$
$\propto$$\nu$$^{-\alpha}$ where $S$ is the flux density at 
frequency $\nu$.

\section{A complete BCG sample.}

To discuss pc scale properties of BCGs in rich clusters of galaxies 
with respect to cluster properties, we need a sample of clusters with no 
selection effect on X-ray and radio properties, and 
distance limited for obvious sensitivity and angular resolution reasons.

Therefore we defined an unbiased complete sample selecting all BCGs in nearby
(Distance Class lower than 3) Abell clusters with a Declination larger
than 0$^{\circ}$. All clusters have been included with no selection on
the cluster conditions (e.g., cooling) and no selection on the BCG
radio power. The Abell catalogue is the most complete and well studied sample 
of rich clusters, and our Declination limit (larger
than 0$^{\circ}$) will not affect our results. This sample is not really
redshift limited, but it is  
representative of
all nearby (z $<$ 0.04) clusters of galaxies.

We present the sample in Table 1. In column 1 we report the name of
the Abell cluster, in column 2 the Distance Class (DC), in
column 3 the redshift $z$ of the cluster, in column 4 we report the
kpc/arcsec conversion for each cluster, in column 5 the name
of the BCG, in column 6 and 7 we give the RA and DEC in J2000
coordinates of the BCG. For sources detected by our new observations
the nuclear position is derived from our image with an estimated
uncertainty $\leq$ 0.1 mas. For undetected sources ((xn) and sources
discussed in the literature), we give in Table 1 the available
coordinates (see notes for the references). In column 8 we
give the angular resolution of maps used to estimate the core
coordinates (see the note of the Tab.1).

Our complete cluster sample is composed of the BCGs of 23 Abell
clusters. The nearest cluster is A262 with redshift = 0.0161 (DC = 1)
and the more distant is A1213 with z= 0.0468 (DC = 2). Some BCGs like
3C\,84 and 3C\,465 are well known and well studied at parsec-scale
with publicly available data and we did not obtain new data for them. In
Table 1, in the last column, we indicate these sources with (!). We
used (*) for binary clusters: where two well defined substructures are
present. In these clusters, we give the name of each BCG for the respective
sub-clusters. This is the case for A1314, A1367 and A2151. We note
also (in the table (**)) that the BCG of A400 is a dumbbell
radiosource with two optical nuclei, 3C\,75A and 3C\,75B. We analyzed
the radio emission associated with both.  In this way, the BCGs that
composed our complete sample are comprised of 27 objects.

\begin{table*}
\caption{The complete BCG sample.}
\begin{center}
\begin{tabular}{ccccccccc}
\hline \hline
Abell Cl.& DC & z   & kpc/arcsec& BCG &RA$_{J2000}$ &  DEC$_{J2000}$& HPBW& Notes\\
\hline \hline
262&0& 0.0161&0.328 &NGC708&  01 52 46.458&  36 09 06.494&0.003$\arcsec$,(v)&  \\ 
 347&1& 0.0187&0.380& NGC910& 02 25 26.77&   41 49 27.3&45$\arcsec$,(x1)& \\    
 400&1& 0.0232&0.468& 3C75A &  02 57 41.650&  06 01 20.717&0.0065$\arcsec$,(v) &(**)\\
&&&&3C75B&02 57 41.563&06 01 36.873&0.006$\arcsec$,(v)&(**)\\   
 407&2& 0.0470&0.922&  UGC2489 & 03 01 51.813&  35 50 19.587&0.005$\arcsec$,(v)&  \\    
 426&0&0.0176&0.358&  3C84&  03 19 48.160&  41 30 42.104&0.0018$\arcsec$,(!1)& \\   
 539&2& 0.0205&0.415&  UGC3274 &  05 16 38.91&  06 25 26.3&45$\arcsec$,(x1)& \\   
 569&1 &0.0196& 0.397& NGC2329&  07 09 08.006&  48 36 55.736&0.0035$\arcsec$,(v)& \\ 
 576& 2&0.0381&0.755&  CGCG261-059&  07 22 06.980&  55 52 30.566&0.004$\arcsec$,(v)&  \\   
 779&1& 0.0226&0.457&  NGC2832&  09 19 46.87 &  33 44 59.6&45$\arcsec$,(x1)&  \\  
1185&2& 0.0304& 0.608 & NGC3550 & 11 10 38.50&   28 46 04.0&5$\arcsec$,(x3)&  \\  
1213&2& 0.0468&0.918 &  4C29.41 &  11 16 34.619&29 15 17.120&0.003$\arcsec$,(v)& \\   
1228&1& 0.0350&0.697 &  IC2738&11 21 23.06 & 34 21 24.0&5$\arcsec$,(x3)&\\   
1314&1& 0.0341& 0.679 &IC708&  11 33 59.222&  49 03 43.428&0.0022$\arcsec$,(v)&(*) \\ 
&&&& IC712  &11 34 49.313 & +49 04 40.54&5$\arcsec$(x3)&(*) \\
1367&1& 0.0215& 0.435 &  NGC3842 &  11 44 02.161& 19 56 59.03&5$\arcsec$,(x5)&(*)\\ 
&&&&3C264 &11 45 05.014&19 36 22.94&0.0035$\arcsec$,(!2)&(*)\\
1656&1& 0.0232& 0.468&NGC4874&  12 59 35.796&  27 57 33.259&0.0027$\arcsec$,(v)&  \\ 
2147&1& 0.0356& 0.708& UGC10143 & 16 02 17.027 &  15 58 28.30&5$\arcsec$,(x3)& \\  
2151&1& 0.0371&0.737& NGC6041  &  16 04 35.794&  17 43 17.551&0.0032$\arcsec$,(v)&(*)   \\ 
&&&& NGC6047&16 05 08.897&   +17 43 54.31&5$\arcsec$,(x3)&(*)\\ 
2152&1& 0.0374&0.742&  UGC10187& 16 05 27.176&    16 26 8.4&5$\arcsec$,(x3)&  \\   
2162&1& 0.0320& 0.639 & NGC6086 &  16 12 35.59 &  29 29 04.8&5$\arcsec$,(x5)&  \\  
2197&1& 0.0303& 0.606 &NGC6173 &  16 29 44.887&40 48 41.880&0.004$\arcsec$, (v)&  \\  
2199&1& 0.0303&0.606 &  3C338&  16 28 38.267&   39 33 04.15&0.0014$\arcsec$,(!3)& \\  
2634&1& 0.0312& 0.624& 3C465&  23 35 58.97& 26 45 16.18&0.0025$\arcsec$ ,(!4)&\\  
2666&1 &0.0265&0.533 &  NGC7768& 23 50 58.549&27 08 50.416& 0.0035$\arcsec$,(v)& \\   
\hline
\hline
\end{tabular}
\end{center}
In column 8 we give the angular resolution of maps used to estimate the core coordinates. We use (v) for the detected objects whose coordinates are taken by the VLBA maps and (!n) for well known radiosources with public VLBI data: (!1)from \citet{tay06}, (!2) from \citet{la97, ba97, la99}, (!3) from \citet{ge07} and (!4) from \citet{ve95}. We use (xn) for undetected radiosources for which coordinates are taken from :(x1), from NVSS; (x2) from NED ; (x3) from FIRST; (x4) from radio data of \citet{fe94}; (x5) from \citet{led95}. In the notes: (*) is for binary clusters for which we give the name of BCGs of the two sub-clusters; (**) indicates a dumbbell radiosource for which we consider the radioemission from both optical nuclei.
\end{table*}

\section{Observations and data reduction.}

To complete our knowledge of the parsec-scale structure of sources in
our sample, we asked and obtained new VLBA observations at 5 GHz for
23 sources. Observations of 10 BCGs were carried out in 2007 June,
while the remaining 13 targets of the complete sample were observed
between 2008 February and March (see Table 2).

All data have been obtained in phase referencing mode. Each source was
observed for $\sim$3 hours. The data were correlated in Socorro,
NM. Postcorrelation processing used the NRAO AIPS package and the
Caltech Difmap packages. We follow the same scheme for the data
reduction of all data sets. We first apply ionospheric and 
Earth Orientation Parameters (EOP) corrections. After this, by running
VLBACALA we correct sampler offsets and apply amplitude calibration.
With VLBAPANG we correct the antenna parallactic angles and with
VLBAMPCL we remove the instrumental delay residuals. All calibrator
data are also globally fringe fitted. After flagging bad data, we
obtain good models for the calibrators, which we use to improve the
amplitude and phase calibration for the entire data set. Final maps
were obtained with DIFMAP by averaging over IFs and in time. After one step
of editing, to obtain a good clean model of the source we applied
multi-iterations of self-calibration in phase to sources with a
correlated flux density $>$ 5 mJy and eventually amplitude
self-calibration to sources with signal to noise ratio $>$ 5.

In Table 2, we report the image parameters for our final maps. For
each source (col.1), we give the epoch of observation (col.2), the
calibrator (col.3), the beam size and P.A.(col. 3 and 4), the noise
(col.5), the intensity peak (col.6) and the total flux density
(col. 7) of final naturally weighted maps.  Typically, the error in
flux density resulting from calibration is $\sim6\%$.  On average, the
resolution of final maps of 2007 data is better (3 $\times$ 1.5 mas) than
2008 images (4$\times$ 2 mas) as the consequence of more antennas (10
instead of 9) available in observations of 2007 compared to that in 2008
(Saint Croix VLBA antenna was not present). The noise level is
$\sim$0.1 mJy/beam, and the detection rate of our new VLBA
observations is 12/23 ($\sim$52$\%$). For detected sources, the
nuclear source position has been estimated with an error $\leq$1
mas. We have to note that the detection rate is likely influenced by
the fact that some BCGs could have an unreliable position derived
from arcsecond images. For these
sources, we imaged VLBI fields as large as a few arcsecs with no time
or/and frequency average allowing us to recover in some cases the nuclear
source. More discussion about the detection rate of our BCG sample is
given in \S5.

\begin{table*}
\caption{Image parameters for naturally weighting maps of new VLBA 5 GHz data.}
\centering
\begin{tabular} {ccccccc}
\hline \hline
Epoch & BCG& Calibrator& HPBW &noise & Peak& Total flux density\\
 yy-mm-dd& & &mas$\times$mas, deg & mJy/beam & mJy/beam& mJy\\
\hline\hline
2007-06-08&NGC2329&J0712+5033 &3.31 x 1.66, 30.20&0.10&62.4&77.1\\
&NGC708&J0156+3914&2.92 x 1.57, -15.90&0.11&3.2&3.2\\
2007-06-14&IC708&J1153+4931&2.24 x 1.81, -4.25&0.10&35.5&44.2\\
&IC712&J1153+4931&2.23 x 1.8, -6.94&0.09&-&-\\
&NGC3842&J1125+2005&3.05 x 1.81, 2.99&0.14&-&-\\
&NGC4874&J1257+3229&2.73 x 1.83, -7.18&0.10&7.37&10.1\\
2007-06-17&NGC6041&J1606+1814&3.20 x 1.70, -6.47&0.09&7.63&7.83\\
&NGC6047&J1606+1814&3.27 x 1.68, -8.09&0.11&-&-\\
&UGC10143&J1606+1814&3.28 x 1.70,6.67&0.12&-&-\\
&UGC10187& J1606+1814&3.31 x 1.7, -7.57&0.11&-&-\\
2008-02-08&NGC910&J0219+4727&3.73 x 2.23, 29.50&0.10&-&-\\
&3C75A&J0249+0619&6.40 x 2.21, 25.4&0.13&20.2&22.0\\
&3C75B&J0249+0619&5.80 x 4.31,19.7 &0.13&37.6&44.5\\
2008-02-09&UGC2489&J0310+3814&4.87 x 2.23, 16.80&0.11&2.6&3.5\\
2008-02-16&NGC7768&J2347+2719&3.51 x 1.40, -7.15&0.09&0.72&1.3\\
&UGC3274&J0517+0648&3.51 x 1.53, 0.28&0.08&-&-\\
2008-02-17&NGC2832&J0919+3324&3.63 x 1.68, 23.10&0.10&-&-\\
2008-02-22&4C29.41&J1103+3014&3.04 x 1.63, -4.56&0.07&36.7&39.2\\
2008-02-28&NGC3550&J1103+3014&3.02 x 1.62,-6.37&0.09&-&-\\
&IC2738&J1127+3620&2.73 x 1.69,- 17.20&0.13&-&-\\
2008-03-16&CGCG261-059&J0742+5444 &4.03 x 1.63, -32.30&0.11&1.9&3.2\\
&NGC6173&J1652+3902&4.07 x 1.94, -3.03&0.11&1.1&1.3\\
&NGC6086&J1634+3203&4.78 x 2.20, -2.54&0.11&-&-\\
\hline\hline
\end{tabular}
\end{table*}

\section{Notes on Individual Sources.}

Here we provide a brief description of all the BCGs in our complete
sample, together with some information on the large scale structure
and cluster conditions. We present contour images for the parsec-scale
radio emission for all resolved sources in our new VLBA observations.

{\bf NGC708 (B2 0149+35) in Abell 262}. This cluster is one of the
most conspicuous condensations in the Pisces-Perseus supercluster. Its
redshift is 0.0161. It is a spiral-rich cluster, characterized by the
presence of a central X-ray source centered on a D/cD
galaxy. \citet{brai94} detect molecular gas and 
suggest that it contains a cooling flow which may contribute to the central X-ray emission.  Observations show a double gaseous component.\\
The BCG has a dust lane almost perpendicular to the direction of its
radio emission \citep{eb85}, which extends along direction
P.A. 70$^{\circ}$ \citep{pa86}.  \citet{bla04} discuss the interaction
between the cooling gas and the radio source which shows at arcsecond
resolution a core with a symmetric structure.  The total 1.4 GHz flux density
is 78 mJy \citep{bla04}. This gives a power at
1.4 GHz of P$_{1.4GHz}$=4.7$\times$10$^{22}$ W/Hz, classifying the
source as a fairly weak,
double-lobed FR I galaxy \citep{fan74}.\\
At parsec-scale, we observed an unresolved structure with total flux
density of 3.2 mJy - about 64$\%$ of the arcsecond core flux
density. The lack of visible jets and the low core radio power are in
agreement with the suggestion that the core is in a low activity
phase and that the offset of the radiative cooling is due to several
outburst episodes \citep{cla09}.
 
{\bf NGC910 in Abell 347}. This cluster at z=0.0187 has been
classified as a small cooling cluster by \citet{whi97}. No powerful
radio galaxy is seen in its central region. The BCG shows in NVSS
images a slightly resolved and faint emission with a flux density of
about 2 mJy, that corresponds to a total power
P$_{1.4GHz}$=1.6$\times$10$^{21}$W/Hz.

Our VLBA observations did not detect a radio source to a 5$\sigma$
limit of $<$ 0.5 mJy/beam.

{\bf 3C\,75 in Abell 400} This cluster has redshift $z=0.0235$. The
central, elliptical-rich population is composed of two subgroups
separated by 700 km s$^{-1}$. These groups are not apparent in the
galaxy distribution projected on the sky.  From this, we can conclude
that a merger is taking place, close to the line-of-sight. Abell 400
does not have a cooling core \citep{ei02}. The central BCG galaxy is
an interesting case of a dumbbell galaxy, with two components separated by
$\sim$7.2 kpc in projection and 453 km s$^{-1}$ in velocity. Twin radio
jets depart from each of the two optical nuclei 3C\,75A and 3C\,75B
\citep{ow85}. The two radio sources have similar total radio power at
1.4 GHz (LogP$_{1.4}$$\sim$24.3 \citet{pa91}), but different nuclear
flux density. The brightest core is 3C\,75B in the north with a core flux
density at 1.4 GHz S$_{core, 1.4}\sim$37.6 mJy compared to 10.5
mJy for 3C\,75A. The large scale radio structure is classified as FRI
\citep{mo93}. The jets are strongly bent by the cluster gas and on
large scales they intertwine and merge (Fig. 1). The strong asymmetry of 
the source to the north and east might suggest motion of both nuclei 
to the southwest relative to the cluster gas.
We observed both nuclei, 3C\,75A and 3C\,75B at mas resolution (see
Fig 1). At parsec-scale, both radio sources are one sided with the main
jet aligned with the kiloparsec structure. 3C\,75A is fainter than
3C\,75B in agreement with the kpc scale behavior. The nucleus of
3C\,75A has a flux density S$\sim$19.9 mJy and the jet has S$\sim$2.14
mJy with an extension of $\sim$4 mas. The core of 3C\,75B has $S$
$\sim$37.6 mJy and the jet has $\sim$7 mJy with an extension of
$\sim$14 mas. The high nuclear flux density of 3C\,75A in the VLBI
image with respect to the arsecond core flux density at 1.4 GHz
suggests a self-absorbed structure and a possible source variability.

{\bf UGC2489 in A407}. No evidence for a cooling core has been found for this
cluster. The BCG appears to be embedded in a diffuse optical halo
within a region of $\sim$60kpc in size \citep{cra99}. In radio images
at arcsecond resolution it shows a complex extended structure with two
lobes and tails oriented East-West. The largest angular size is
$\sim$250 arcsec and the total flux density at 20 cm S$_{t,20}$
$\sim$728 mJy.  The core has S$_{c,20}$ $\leq$10 mJy, the East
lobe+tail is $\sim$110 arcsec extended with S$_{E,20}$$\sim$305 mJy
and the West one is brighter (S$_{W, 20}$$\sim$ 416 mJy) and larger
($\sim$ 123 arcsec) \citep{bo93}.

At mas scale (fig. 2), the source appears one-sided with a core flux
density $\sim$2.6 mJy. The jet direction is NW-SE with an extension of
6 mas and flux density $\sim$ 0.9 mJy.

{\bf 3C\,84 in A426} The Perseus Cluster, A426, is the most X-ray
luminous cluster in the nearby universe, and the prototypical “cooling
core” cluster. Shocks and ripples are clearly evident in the deep
Chandra image of Perseus \citep{fa05, fa06}, and could provide steady
heating of the center of the cluster \citep{fa06}. In Perseus the AGN
manifests itself directly as a bright radio source known as Perseus A
or 3C\,84, associated with the early-type cD galaxy NGC1275.  3C\,84
is one of the brightest compact radio sources in the sky and has been
studied in some detail \citep{ver94, tav96, sil98, wal00}.  The radio
source 3C\,84 possesses well known jets which have been documented on
a variety of scales \citep{pe90, dha98, sil98, wal00}.  
\citet{tay06} detect linear polarization from the bright
jet component S1 in 3C\,84 at 5, 8, 15 and 22 GHz at a level of 0.8 to
7.5$\%$ increasing with frequency. Furthermore, there is some
suggestion at 8.4 GHz and above that the polarization is extended. The
detection of core polarization is less than 0.1$\%$ for all
frequencies except for 22 GHz for which it is less than 0.2$\%$.

The radio morphology is quite complex \citep{asa09, lis01, agu05}
exhibiting a core with two opposite radio-jets, with the southern jet
consisting of components moving down a diffuse jet and finally
expanding into an amorphous component at 12mas.  \citet{kri92} showed
that the inner jet components move at 0.1c and that after a major
bend the jet speed accelerates.\\
The jet morphology has been imaged on a variety of scales. At high frequency (and resolution) we compared the VLBA image at 43 GHz \citep{lis01} with the 86 GHz image by \citet{lee08}. Despite different epochs and resolutions we tentatively identify the core with component $``$E$''$ in \citet{lis01} being the only self-absorbed structure (inverted spectrum) between 43 and 86 GHz. Therefore the source appears initially one-sided, and later becomes two-sided.\\
At lower frequency and resolution, there is a clear equivalence
between the parsec-scale structure and jets seen on
kiloparsec-scale \citep{pe90}. Looking through VLBA images from 22 GHz to 5 GHz, the two-sided structure starts out appearing more symmetric and then at lower frequencies it becomes assymetric. \citet{wal98, wal00} explained this as effects of free-free absorption of the counter-jet from the surrounding torus.  VLBA data at 15 GHz \citep{lis09}
measure proper motion in the jet and counterjet and estimate
$\gamma$=0.6 and $\theta$=11$^{\circ}$.  In VLBA data at 5 GHz
\citep{tay06} the largest source angular size is $\sim$ 35 mas
($\sim$12 pc) , the core flux density is $\sim$ 3.1 Jy (P $\sim$ 2.1
$\times$ 10$^{24}$ W/Hz) and the total flux density is 23.3 Jy (P
$\sim$ 1.6 $\times$ 10$^{25}$ W/Hz).

{\bf UGC3274 in A539}. This cluster is at z= 0.0205; its internal
structure is not clearly understood. \citet{gir97} found that it is
composed by two systems, both very extended, separated in velocity by
over 4000 km/sec, but spatially overlapped. The BCG in NVSS image is
not detected ($\leq$ 1.4 mJy) which corresponds to an upper limit on
the total power LogP$\leq$21.1(W/Hz).
Our VLBA observations did not detect a radio source to a 5$\sigma$
limit of $<$ 0.4 mJy/beam.

{\bf NGC2329 in Abell 569}. No cooling core has been found in this
cluster. The galaxy has an exceptionally blue V-I ($\sim$1.20) color
\citep{ve99}.  The nucleus is clearly bluer than its surroundings. The
nuclear emission is slightly resolved and shows a small extension
roughly in the direction of the dust protuberance. The galaxy has a
small inclined central dust disk.

NGC2329 has been associated with a wide-angle tail (WAT) radio source
with two tails extended for about 4 arcmin. The total power at 1.4
GHz is LogP=23.48 W/Hz and the core power at 5 GHz is LogP=22.52 W/Hz
\citep{fe85}. The extended radio morphology suggests merging.

The VLBA observations at 1.6 GHz of \citet{xu00} detect a one-sided
morphology with peak flux density $\sim$49.7 mJy/beam and total
flux density $\sim$59.1 mJy. The core-jet is elongated in the
direction corresponding to the northeastern radio tail. We confirm
this morphology in our data. In our images (fig. 3), the core has
S$\sim$66.8 mJy, the jet has S$\sim$ 10.3 mJy with an extension of
$\sim$ 15 mas. The total spectral index derived from 1.6 and 5 GHz
data is $\alpha$$\sim$ -0.2.

{\bf CGCG 261-059 in A576}: This cluster is not relaxed. The rise in
velocity dispersion of emission-line population towards the cluster
core indicates that this population is dynamically aware of the
cluster and probably falling into the cluster for the first time. The
observations seem to be in favour of the model for which the core of
A576 contains the remnants of a lower mass subcluster \citep{moh96}.
The BCG is slightly extended in the NVSS image and it shows a mJy
level emission. In the FIRST image it appears pointlike with a flux
density $\sim$ 3.33 mJy that corresponds to total power LogP=22.04
(W/Hz).

In our parsec-scale map (fig. 4), the source is one sided with a core
flux density of 1.87 mJy. The jet has a flux density of 1.37 mJy, and 
extends for 7 mas in P.A $\sim$-20$^{\circ}$.

{\bf NGC 2832 in Abell 779}. This is a poor cluster, with no evidence
of a cooling core from X-ray data \citep{whi97}. In high-resolution
ROSAT HRI soft X-ray observations, NGC 2832 shows a strong central
emission peak, with some diffuse emission around it. It does not
possess an extended cD envelope \citep{mal81}. A nearby satellite
galaxy, NGC 2831, appears to be tidally interacting with the BCG
despite a relative velocity difference of 1692 km s$^{-1}$
\citep{lau88}. No optical emission lines are present in NGC 2832, and
it lacks a cooling core \citep{mcn92}.

The radio emission associated with the BCG is at the mJy level, unresolved
in FIRST images with a flux density $\sim$ 2.36 mJy that corresponds
to a total power LogP=21.43(W/Hz). Our VLBA observations did not
detect a radio source to a 5$\sigma$ limit of $<$ 0.5 mJy/beam.

{\bf NGC3550 in Abell 1185}. This cluster is at z=0.0304 with no
evidence for a cooling core\citep{whi97}. It is undetected in NVSS and
FIRST images.  In our image, we did not detect any radio emission
$\geq$ 0.45 mJy/beam (5$\sigma$).

{\bf 4C 29.41 in Abell 1213}. This cluster is underluminous in the
X-rays, just marginally detected by the Einstein and ROSAT
satellites. The X-ray images and optical distribution suggest a non
relaxed structure \citep{jon99}. At kiloparsec-scale, this radio source
is part of a dumbell galaxy \citep{tru97}. It is an example of double
and symmetric radio source with a bright one-sided jet. It is
classified as a FRI radiogalaxy with total power at 1.4 GHz LogP$_{1.4
  GHz, t}$$\sim$ 25.35 and core flux density at 5 GHz of S$_{5 GHz,
  core}\sim$ 41 mJy. Its linear size is about 81 kpc \citep{ow97}.

On the mas scale (fig.5), it shows one-sided morphology with the jet
aligned with the kiloparsec structure. The core flux density is
$\sim$36.24 mJy, while the jet has $\sim$ 2.5 mJy and is visible out
to $\sim$ 12 mas.

{\bf IC 2738 in A1228}. This cluster is at z= 0.035. No cooling core has
been detected \citep{whi97}. Its BCG is undetected in NVSS maps with a
power LogP$<$21.14 (W/Hz). In our image, no detection is present at
the 5$\sigma$ level (0.65 mJy/beam).

{\bf IC 712 and IC 708 in Abell 1314}. This cluster is a binary
cluster with two main condensations visible in the X-ray images.
This cluster shows very clumpy, elongated X-ray emission, and a strong
X-ray centroid shift \citep{bli98}. This elongation is also present in
the galaxy distribution, which shows a definite ellipticity, and is
oriented mainly E-W \citep{fli95}. IC\,712 is the BCG of the main
condensation and it shows a WAT structure with a very small linear
size ($\sim$ 4.6 kpc) with a total flux density at 1.4 GHz of 26.3 mJy
\citep{gio94}. IC 708 is the BCG of the second condensation. It also
shows a WAT structure \citep{va79} with a larger size
($\sim$4.5 arcsec), total flux density at 1.4 GHZ S$_{1.4
  GHz,t}$$\sim$ 430 mJy and core flux density at 6 cm S$_{5GHz,
  core}$$\sim$110 mJy.

In our maps, IC 712 is undetected at the 5$\sigma$ level ($<$0.45 mJy/
beam). IC 708 (fig.6) shows a one-sided structure with a core flux
density $S \sim$38.8 mJy. The jet is aligned with the kiloparsec
structure, has S$\sim$ 2.48 mJy and an extension of $\sim$ 16 mas
from the core.

{\bf 3C\,264 and NGC3842 in Abell 1367}. A 1367 is an X-ray-faint and
nearby (z=0.0215) galaxy cluster. It has a secondary peak of the X-ray
brightness \citep{don98} located approximately 19' offset from the
primary brightness peak. This cluster is peculiar in the existence of
extended trails of radio emission behind three irregular galaxies in
its periphery \citep{ga87}.  It is a complex cluster currently forming
at the intersection of two filaments. NCG3842 and NGC 3862 are the two
brightest cluster galaxies.

NGC3842 is a cD galaxy studied in radio band by \citet{fe94}. From
high resolution VLA radio images, NGC3842 shows a small size WAT
structure with angular size of $\sim$55 arcsec and a flux density of
10.5 mJy corresponding to a total power at 1.4GHz
LogP$_{t,1.4GHz}$=21.72 W/Hz. In our images it appears undetected above
5$\sigma$ ($<$ 0.7 mJy/beam).

NGC3862 is associated with the strong radio source 3C\,264.  In VLA
images, it has a total power LogP=24.97 (W/Hz).  3C\,264 has a head-tailed
morphology at kiloparsec-scales, a prominent core and a wiggling jet
extending toward the northeast that ends in a blob of emission at 28
arcsec (11.5 kpc) from the core \citep{la97}. There is evidence of
counterjet emission in the southwest direction from the core. Both the
jet and counterjet are embedded in a vast and diffuse region of low
surface brightness emission which seems to have beeen dragged toward the
north, possibly revealing the existence of a high density intracluster
medium.  Simultaneous EVN and MERLIN observations at 5.0 GHz by
\citep{la99, ba97} showed for the first time the detailed structure of
3C\,264 at sub-kiloparsec-scales. It consists of a one-sided jet with
evident variations in its morphological properties with distance: i)
the strong core and innermost jet (0-10 pc); ii) a well-collimated and
narrow region (10-100 pc); iii) a region with strong widening, kinks
and filaments (100-300 pc); iv) a faint and narrow region after a jet
deflection (300-400 pc from the core).  In particular, in the EVN map at 5
GHz \citep{la99}, the source shows an unresolved core with a peak flux
density $\sim$ 0.126 Jy/beam and a smooth one-sided jet extending up
to 25 mas from the core along P.A.$\sim$27$^{\circ}$.

{\bf NGC4874 in Abell 1656}. This is one of the two dominant members of
the Coma Cluster (Abell 1656) and considered the Coma BCG. It is a cD
galaxy that shows a small size WAT structure. Arcsecond scale
properties are discussed in \citet{fe87}, where a possible precessing
beam model is discussed. On kiloparsec-scales, it has a total flux density at
1.4 GHz of 190 mJy and the core flux density at 6 cm is 1.1 mJy. The
radio emission is completely embedded in the optical galaxy. In fact,
the radio linear size is 7 kpc.
In the VLA 1.4 GHz image, a gap of radio emission is present between the core and the SW lobe, while a faint jet connecting the core and the NE lobe is detected. At 5 GHz the jets are detected after gaps of $\sim$3 arcsec on both sides of the core.

In our VLBA image (fig. 7), the source shows a one-sided structure
with a total flux density of 10.1 mJy. The core flux density is 7.27
mJy. The jet has flux density $\sim$2.83 mJy and it is extended in
direction SW (P.A.$\sim$ - 15.4$^{\circ}$) with an angular size of
$\sim$ 3 mas.

{\bf UGC10143 in Abell 2147.} This cluster shows in the X-ray images,
extended elongated emission suggestive of a merger.
\citep{flin06}. The BCG is a cD galaxy with an elongated radio
emission apparent in the NVSS image. In the high resolution image from 
FIRST, it shows a WAT morphology with angular size $\sim$20 arcsec and
total flux density $\sim$8.03 mJy. At mas scale, we did not detect a
radio source: the 5$\sigma$ limit is $<$ 0.6 mJy/beam.

{\bf NGC 6041A and NGC 6047 in Abell 2151}. This Abell cluster is also
known as the Hercules cluster. It is a highly structured cluster
despite an apparently regular velocity distribution of the main
field. This cluster should be considered as an ongoing cluster
merging\citep{gir97}. This cluster shows two peaks in the X-ray
emission \citep{bi95}. There is only marginal evidence of soft excess
emission in the brightest of the two peaks \citep{bo02}.

NGC 6041A is the BCG of the first main condensation while NGC 6047 is
the brightest galaxy of the second condensation. In the NVSS image,
NGC 6041A shows a WAT structure with angular size $\sim$2.5 arcmin and
total flux density $\sim$ 11.6 mJy. In our VLBA map, it appears
unresolved with a total flux density $\sim$7.83 mJy.

NGC 6047 is classified as an E/S0 and it has a bizarre optical morphology
which suggests that it may be a recent merger which has so severely
disrupted the dynamics of an HI disk that the gas has lost its
kinematic coherence \citep{di97}. At kiloparsec-scale, NGC 6047 shows
an extended FRI structure with a two-sided jet emission
\citep{fe88}. The northern jet is brighter (flux density at 5 GHz
$\sim$13.6 mJy) and more collimated. The total flux density at 1.4 GHz
is $\sim$ 728 mJy and the core flux density at 5 GHz is $\sim$8.3
mJy. In our maps, it appears undetected above 5$\sigma$= 0.55
mJy/beam.

{\bf UGC10187 in A2152.} A2152 is one of the major condensations in
the Hercules supercluster together with A2147 and A2151. Its BCG is
UGC10187, which is also the brightest of a galaxy pair. In the NVSS
this BCG shows an extended ($\sim$ 4 arcmin in NVSS) tailed radio
structure with total flux density $\sim$60.3 mJy. In the FIRST image,
a gap of $\sim$5 arcsec between the two tails is present and the
nuclear source is not visible. In our VLBI image, no detection was
found with a 5$\sigma$ limit $<$ 0.55 mJy/beam.

{\bf NGC6086 in Abell 2162.} This Abell cluster is a low X-ray
luminosity cluster \citep{bu94}. Its BCG is a bright cD galaxy which
hosts a double-lobed radio source.  In the NVSS
image, the total flux density is $\sim$108.7 mJy and its angular size
is $\sim$ 3.5 arcmin.  In the FIRST image, the radio emission is partly
resolved out (only $\sim$60 mJy) and extended over $\sim$2.5
arcmin. No defined structure is observable. The radio spectrum and
morphology suggest that it is a relic galaxy where the core radio
activity stopped some time ago. Our VLBA observations did not detect a
radio source to a 5$\sigma$ limit of $<$0.55 mJy.

{\bf NGC6173 in A2197.} This cluster is one of a few clusters which
show an unusual systematic alignment among individual cluster members
and it is possible that is linked with A2199 via supercluster
structure to the Hercules region \citep{gre84}.

The BCG shows mJy level emission ($\sim$7 mJy) unresolved in FIRST
images.  At mas scale, it appears (fig.8) one-sided with a total flux
density in our VLBA maps of 3.24 mJy. In particular, the core has a
flux density of 1.87 mJy and the jet is extended to the east
with an angular size $\sim$ 10 mas and flux density $\sim$ 1.37 mJy.

{\bf 3C\,338 in Abell 2199.} A2199 (z = 0.0304) has a prototypical
cooling core. In fact, a Chandra observation of this cluster has
detected a point-like source coinciding with the center of the galaxy
NGC 6166 \citep{mat01} that is a multiple nuclei cD galaxy. This
galaxy hosts the relatively powerful radio source 3C\,338, which emits
a total power at 330 MHz of LogP$\sim$25.64 W/Hz. This radio source
has been known for a few decades to have an unusual structure on both
large and small scales \citep{fe93, gio98}. It is classified as a FR I
radio source and shows central optical [O III] line emission
\citep{fi95}.  On kiloparsec-scales 3C\,338 has two symmetric extended
radio lobes, characterized by a steep spectrum ($\alpha$$\sim$-1.7)
and misaligned with the central emission. The two radio lobes are
connected by a bright filamentary structure. Polarimetric observations
by \citet{ge94} revealed strong rotation measure gradients across most
of the extended emission and inferred the presence of cluster magnetic
fields. Both the steep radio spectrum and strong filamentary emission
may be the result of interactions with the dense intracluster medium
\citep{ge07}.

On parsec-scales, 3C\,338 has a compact radio core with two short
($\sim$10 pc), symmetric jets. In VLBA maps at 5 GHz \citep{fe93}, the
parsec-scale structure shows a central dominant feature (the core
emission) with flux density $\sim$ 92 mJy/beam and two symmetric
jets. The total flux density is 133 mJy. The eastern jet shows a
couple of low-brightness regions in its center suggesting it could be
limb-brightened. The orientation of this structure appears to be very
constant in time despite the complex dynamic conditions present in the
3C\,338 central regions.  Comparing maps obtained at different epochs
\citep{ge07}, a change in the parsec-scale morphology is clearly evident,
and it is probably correlated with the arcsecond core flux density
variability. The structural changes suggest the presence of proper
motion with $\beta$$\sim$0.4 h$^{-1}$ on both sides of the core
\citep{ge07}. These properties suggest that the extended emission in
3C\,338 is a relic structure not related to the present nuclear
activity \citep{gio98}.

{\bf 3C\,465 in Abell 2634}. A2634 (z=0.0322) does not show evidence
of a cooling core.  Its BCG is 3C\,465 that is associated with the
giant D galaxy NGC 7720. The optical host galaxy exhibits distorted
isophotes with two gravitationally bound nuclei surrounded by a common
envelope \citep{ve95}. Weak broad H$_{\alpha}$ and high-ionization
emission-lines suggest the presence of faint activity in the optical
nucleus.  In the radio band, this source has a WAT morphology. Its
total power at 408 MHz is LogP=25.3. The radio jet, the spots and the
low-brightness tails are strongly polarized.  On parsec-scales the
source is one-sided with the jet on the same side as the main
kiloparsec-scale structure. At 5 GHz, from VLBA data \citep{ve95}, the
core flux density is 168 mJy and the total flux density is 237 mJy.

{\bf NGC7768 in Abell 2666.} \citet{sco95} did not find any evidence
of substructures in the central cluster region. Its BCG is a cD galaxy
in the center of the cluster. It contains a dusty nuclear disk
approximately aligned with the major axis of the galaxy \citep{gri94}
and radio emission at the mJy level ($\sim$2 mJy that corresponds to
LogP=21.50 W/Hz).

In our VLBA data (fig. 9), it shows a one sided structure with a core
flux density $S \sim$0.72 mJy and total flux density $\sim$1.31
mJy. The jet is oriented in direction NW-SE, it has $S \sim$ 0.59 mJy
and an extension of $\sim$ 4 mas from the core.

\section{Parsec-scale morphology.}

Our complete sample is composed of 27 BCGs, 23 have been observed for
the first time with VLBA by us and for the remaining objects radio
information at mas scale are available from literature. The detection
rate of our new VLBA observations is 52$\%$ and, adding literature
parsec-scale information of the well studied radiosources, the total
detection rate at mas resolution of our complete sample is
59$\%$(16/27). In particular, 45$\%$(12/27) are one-sided,
7$\%$(2/27) have a two-sided morphology, 7$\%$(2/27) show point-like
structure, and 41$\%$(11/27) are undetected.
Table 3 presents parsec-scale and arcsecond radio properties for all
our BCGs (complete sample and extended sample, see section 7).

\subsection{Two-sided morphology.} In our sample, two sources,
3C\,84 and 3C\,338, show two-sided structure. For both radio sources,
the most supported explanation suggests that their properties can be
explained if they consist of two oppositely directed, symmetric, and
mildly relativistic jets at a modest angle to the line of sight
(\citet{walk94, dha98} for 3C\,84 and \citet{gio98, ge07} for 3C\,338).

Both 3C\,84 and 3C\,338 lie at the center of cool core clusters. The
Perseus cluster, Abell 426, in particular is the most X-ray luminous
cluster in the nearby universe, and the prototypical 'cooling core'
cluster. Moreover, these two sources show evidence of a restarted
activity. In 3C\,338 there is evidence of restarted radio activity
with an extended relic emission and a small-scale young structures
\citep{gio98}. Features suggesting the recurrent jet activity of the
central engine are present also in 3C\,84. At low resolution
(frequencies below 5 GHz), the southern jet extends continuously from
the core out to $\sim$100 mas \citep{tay96}, indicating that 3C 84
had previous outbursts. Also, there are multiple lobe-like structures
at arcminute scale \citep{pe90}. Moreover, the synchrotron age of
inner lobe \citep{na09} and the observed inner proper motions
\citep{asa09} is consistent with the scenario that the inner lobe
formed by the 1959 outburst \citep{nev95}.

As discussed in previous section 3C 84 appears one-sided at 86 GHz in the inner
2.5-3 mas ($\leq$ 1 pc) and symmetric (two sided) at $\sim$ 4-5 mas
($\geq$1.5 pc) from the core as evident in the 22 GHz images by
\citet{tay06}.  The jet/counterjet ratio is $>$35 at 43 GHz at 0.5 mas
($\sim$0.15 pc) and $\sim$1 at 5 mas ($\sim$1.5 pc) suggesting a
strong jet deceleration. This result cannot be due to free-free absorption effect \citep{wal00} since the one sideness is visible in the high resolution high frequency image. We note that a similar structure can be
detected only here because of the good linear resolution due to the low redshift of 3C 84.

\subsection{One-sided morphology.} In our complete sample 12 BCGs 
show one-sided parsec-scale structures. The percentage (45$\%$) is in
agreement with the percentage of one-sided FRI radio galaxies found in
a complete sample of radio galaxies \citep{liu09b}: 23/51 (45$\%$). We
note that all 12 BCGs lie in non-cool core clusters.

The values of the jet/counter-jet ratios for the sources of our
complete sample are given in Table 3. We used these values together
with the core dominance information to estimate the angle and jet
velocity \citep{gio01}. Results are uncertain but consistent with an
asymmetry because of Doppler boosting effects.

\begin{table*}[htp]
\caption{{\bf Parameters for BCGs.} We give the name of the BCG (col.1), the parsec-scale mophology (col.2 : core (c), one sided (1s) or two sided (2s) jet structure), the jet/counterjet surface brightness ratio (col.3), $\beta$co$\theta$ (col.4 ), the arcsecond core flux density S$_{c, 5}$ at 5 Ghz (col.5), the arcsecond core power LogP$_{c, 5 }$ at 5 GHz (col.6), the total arcsecond flux density S$_{t, 408}$ (col.7) and the total arcsecond power LogP$_{t, 408 }$ (col.8) at 408 MHz, the total VLBI flux density and arcsecond core flux density ratio S$_{VLBI}$/S$_{c, 5}$ at 5 GHz (col. 9), the core dominance (see section 5.2) (col.10). The last column is for the notes. }
\begin{center}
\label{tab:id}
\scriptsize
\tabcolsep2mm
\begin{tabular}{ccccccccccc}
\hline
\hline
BCG&pc &j/cj & $\beta$cos$\theta$&S$_{c,5}$&LogP$_{c,5}$&S$_{t,408}$&LoP$_{t,408}$&S$_{VLBI}$/&core&notes\\
&structure&ratio&&mJy&W/Hz&mJy&W/Hz&S$_{c, 5GHz}$($\%$)&dominance&\\
\hline \hline
NGC708& c   &-&-&5&21.47&364&23.33&60&0.26&5\\ 
NGC910& n.d.&-&2&-&-&-&-&$<$25&-&5\\
3C75A&1s&$\geq$6.2&$\geq$0.35&6&21.87&190&23.37&100&0.60&1\\ 
3C75B&1s&$\geq4.6$&$\geq$0.3&20&22.39&190&23.37&100&2&1\\ 
3C84&2s&2&0.14&28.17$\times$10$^{3}$&25.29&51.68$\times$10$^{3}$&25.56&80&70.79&2\\ 
UGC3274&n.d.&-&-&$<$1.4&-&-&-&-&-&5\\
UGC2489&1s&$\geq$2.8&$\geq$0.2&4&22.32&2.83$\times$10$^{3}$&25.17&90&0.13&3\\
NGC2329&1s&$\geq$4.5&$\geq$0.29&160&23.14&1300&24.05&50&4.3&4\\
CGCG261-059&1s&$\geq$1.8&$\geq$0.12&1.5&21.71&6.54&22.34&100&1.8&5\\ 
NGC2832&n.d.&-&-&2.4&-&-&-&$<$21&-&5\\
NGC3550&n.d.&-&-&$<$1.4&-&-&-&-&-&5\\
4C29.41&1s&$\geq$3.8&$\geq$0.26&41&23.33&135.2&23.84&100&8.71&6\\
IC2738&n.d.&-&-&-&-&-&-&-&-&5\\
IC708&1s&$\geq$6.5&$\geq$0.36&110&23.47&901.1&24.39&40&5.62&7\\ 
IC712&n.d.&-&-&14.02&22.58&48.72&23.12&$<$3&4.37&11\\
NGC3842&n.d.&-&-&9&22.98&101.88&23.03&$<$8&0.36&11\\
3C264&1s&$\geq$6&$\geq$0.34&200&23.32&17$\times$10$^{3}$&25.25&100&1.17&8\\ 
NGC4874&1s&$\geq$1.6&$\geq$0.09&1.1&21.13&351.96&23.63&100&0.08&9\\ 
UGC10143&n.d.&-&-&$<$2&$<$21.77&14.87&22.64&-&$<$1.35&5\\
NGC6041A&c&-&-&$\leq$0.9&$\leq$21.92&21.5&22.84&100&1.45&5\\ 
NGC6047&n.d.&-&-&8.3&22.42&1960&24.80&$<$7&0.28&12\\
UGC10187&n.d.&-&-&$<$2&$<$21.81&111.7&23.56&-&$<$0.41&5\\
NGC6086&n.d.&-&-&$<$1&$<$21.37&201.35&23.68&-&$<$0.12&5\\
NGC6173&1s&$\geq$1.4&$\geq$0.07&3.7&21.89&12.97&22.44&40&2.40&5\\
3C338&2s&2.2&0.16&480&24.01&18.12$\times$10$^{3}$&25.59&30&3.47&10\\ 
3C465&1s&$\geq$20&$\geq$0.54&246&23.74&10.38$\times$10$^{3}$&25.37&100&2.57&8\\ 
NGC7768&1s&$\geq$1.1&$\geq$0.02&0.74&21.08&2.6&21.62&100&1.17&5\\ 
\hline
B2 0836+29II&1s&$\geq$20&$\geq$0.54&131&24.31&1139&25.24&100&11.20&13\\
Hydra A&2s&1.2&0.04&168&24.08&132$\times$10$^{3}$&26.98&100&0.58&14\\
4C 26.42&2s&1.4&0.07&53&23.71&3153&25.48&80&2.04&15\\
3C 317&2s&1.5&0.08&310&23.93&132$\times$10$^{3}$&26.56&100&0.72&16\\
B2151+174&2s&4&0.27&164&25.40&538&25.92&100&53&17\\
PKS 2322-123&2s&1.2&0.04&59.3&24.0&7.2$\times$10$^{3}$&26.11&80&1.66&18\\
PKS 1246-410&1s&$\geq$9&0.34&64.4&22.27&2463&23.86&30&0.76&19\\
\hline
\end{tabular}
\end{center}
Notes: S$_{t, 408}$: 1) From NVSS with $\alpha\sim$0.5, 2) from \citet{pa91} with $\alpha\sim$0.5; 3) from \citet{pe90} where $\alpha\sim$0.5, 4) from \citet{bo93}, 5)from \citet{fe85},  6) from \citet{ow97} with $\alpha\sim0.5$, 7) from \citet{va79} where $\alpha\sim$0.6, 8) from \citet{fe94},  9) from \citet{gio01} and NED informations with $\alpha\sim$0.5, 10) from \citet{fe87} with $\alpha\sim$0.5, 11) from \citet{ow97}, 12)from \citet{tay02} and NED information,  13) from \citet{ve95}, ), 14) from \citet{tay90, tay96}, 15) from \citet{liu09}, 16) from \citet{ve04}, 17) from \citet{au06} and from NED with $\alpha\sim$0.5 (408 MHz) and $\alpha\sim$0 (5 GHz), 18) from \citet{tay99} and NED information and 19) from \citet{tas06} assuming $\alpha$=0.5. S$_{c}$ is for the arcsecond core flux density at 5 GHz and S$_{VLBI}$ is a correlated flux density at 5 GHz in our VLBI data.
\end{table*}

Moreover, we note that in all resolved BCG sample sources the parsec
scale jet is aligned with the arcsecond structure indicating that no
complex strong change in the angular momentum of the accreted gas, and
no restarted activity with different inclination of the accretion disk
and or central BH precession occur.

\subsection{Unresolved sources.} At milliarcsecond resolution, two
radio sources of our complete sample appear unresolved: NGC708 in the
cooling cluster A262 and NGC6041A in the merging cluster A2151.
We note that, for NGC 708, the core dominance, defined as the ratio
between the observed and the estimated core radio power according to
the relation given in \citet{gio01} (see also \citet{liu09b}) is very
low (0.25) suggesting that the nuclear activity is in a low phase and
it is for this reason that the jets are not visible.
NGC6041A is a faint source (LogP $\sim$22.57 (W/Hz) at 1.4 GHz) and the
parsec-scale jets -if any- are probably too weak to be detected with
the present sensitivity.

\subsection{Undetected sources.} In our new VLBA observations, 11 BCGs are
undetected below 5$\sigma$. This percentage is high but still
significantly lower that in non-BCG ellipticals. The nature of the 
undetected BCGs is varied and they can be grouped as follws:
\begin{center}
 \begin{itemize}
\item radio quiet sources (5/11): this is the case for UGC3274, NGC2832, NGC3550, IC2738 and NGC910 that do not show any radio emission at arcsecond resolution;
  \item radioquiet core (4/11) in a radio galaxy: in these sources the central AGN was active in the past, but is radio quiet at the time of observations. This is the case for NGC6086, a candidate relic radio galaxy, for the WAT NGC3842 where the arcsecond core is very faint ($\sim$0.26 mJy), for UGC10143 and for UGC10187 where the nuclear source is not detected by the VLA radio images. As expected in these sources the core dominance is low;
\item peculiar sources (2/11): NCG 6047 and IC712. In VLA maps, these sources show radio emission from the core but they appear undetected in our VLBA maps. 
This could be due to an extreme variability of the core emission. Alternatively, these sources could have a complex structure on scales between these allowed by VLBA and VLA with a pc scale low surface brighteness that we are not able to map with our VLBA data. In these cases, more sensitive VLBA observations, EVLA observations at high frequencies, or e-Merlin observations will be important to properly study these structures.
\end{itemize}
\end{center}
 
\section{Results for the complete sample.}

 We note that in our complete sample we have 23 clusters 
of galaxies, and only 
5 of them (22\%) have been defined cooling core clusters. This percentage is
lower than values found in literature: e.g. \citet{hud09} found that 44\% of 
galaxy clusters have strong cool cores. However we note that their statistic
is based on X-ray flux 
limited samples, while our sample has no selection effects on the X-ray 
luminosity. If we apply the same constraints of \citet{hud09} to our sample 
we will have only
9 clusters, and 3 of them (33\%) with a strong cool core, in agreement
(note the small numbers) with literature data.

\subsection{Statistical considerations.}

\begin{itemize}
\item We compare the total flux at VLBA scales with the core arcsecond
  flux density (Tab. 3). Over all data, among 16 detected sources, we
  find that 11/16 (70$\%$) have a correlated flux density larger than
  80$\%$ of the arcsecond flux density. This means that in these
  sources we imaged most of the mas scale structure and so we can
  properly connect the parsec to the kiloparsec structures. In contrast,
  for 5/16 (30$\%$), a significant fraction of the arcsecond core flux
  density is missing in the VLBA images. This suggests variability or
  the presence of significant structures between $\sim$ 10 mas and 1
  arcsecond that the VLBA can miss due to the lack of short
  baselines. To properly study these structures, future observations
  with the EVLA at high frequency or with the e-MERLIN array will be
  necessary.
\item  We derive the distribution of the total radio luminosities at
  1.4 GHz obtained from NVSS \citep{con98} and FIRST \citep{bec95} and
  the RLF (Radio Luminosity Function) of the radio loud sources in our
  complete sample. We compare our sample properties with the results
  of \citet{be07} for their complete sample of BCGs. Because of the
  lack of spectroscopic information of our BCGs, we are not able to
  distinguish between radio emission due to star formation and due to
  AGN activity. Following \citet{be07}, we compared only sources
  with LogL$\geq$22.3 and we assumed that the detected radio emission
  of our BCGs is from central AGN.

We found that in our complete sample the probability
  of BCGs to be radio loud with LogL$\geq$22.3 is $\sim$87$\%$. This is
  consistent with the results of \citet{be07}. We consider this result
  as evidence that our small sample is representative of the general
  properties of BCGs.

\item Among the sources of our complete sample, there are two cases,
  3C\,84 and 3C\,338, where there is evidence for restarted activity
  in the radio emission associated with the BCG. Both 3C\,84 and
  3C\,338 lie in Abell clusters that show a presence of a cool core.

  In the case of 3C 84, \citet{na09} and \citet{agu05} used multifrequency VLBA
  observations to constrain the timescale of the restarted
  activity. In particular, \citet{agu05} derived a kinematic age and
  \citet{na09} estimated the synchrotron age of the radiosource. Both
  authors found a resulting age which is consistent with the scenario
  that the inner $\sim$15 mas feature is formed by the recent outburst
  in 1959. At low frequencies, the southern jet extends continuously
  from the core out to $\sim$ 100 mas, suggesting that 3C\,84 has had
  multiples outbursts. Also there are multiple lobe-like structures at
  arcminute scales that suggest a recurrent jet activity of the
  central engine.

  For 3C\,338, due to the peculiar morphology and arcsecond core flux
  variability, \citet{gio98} suggest that the extended emission is
  older and unrelated to the present nuclear activity (see also
  \citet{ge07}).

\end{itemize}

\section{The extended sample.}

Looking at X-ray emission of the Abell clusters in our complete
sample, clusters are divided in cool core and non cool clusters
(Tab.4). However, at parsec-scale, we note that the two-sided
structure (3C\,84 and 3C\,338) are only found in cool core clusters (A
426 and A 2199) and one-sided sources are all observed in non cool
core clusters. A comparison between BCGs in cooling and non-cooling
clusters suggests a difference in the properties of the parsec-scale
structures, but numbers are too small to properly discuss it. To
improve our statistics, we performed a search in the literature and
archive data looking for VLBI data of BCG in Abell clusters that have
detailed information about the cluster dynamic (X-ray emission) and
radio emission at mas and arcsec resolution. In order to obtain an
extended BCG list, we added to our complete sample the following
sources:
\begin{center}
 \begin{itemize}
  \item {\bf B2 0836+29II in A690} \citep{odo90, gio05};
\item {\bf Hydra A in A780} \citep{tay96, wi07};
\item {\bf 4C 26.42 in A1795} \citep{liu09, sal04};
\item {\bf 3C317 in A2052} \citep{ve04};
\item {\bf B2151+174 in A2390} \citep{au06};
\item {\bf PKS 2322-123 in A2597} \citep{mo05, tay99}
\item {\bf PKS 1246-410 (NGC4696) in A3526} \citep{tas06}
 \end{itemize}
\end{center}

  In Tab.4 we provide our results concerning the morphology for the extended sample. We report the Abell cluster of BCG (col. 1), X-ray cluster properties (col. 2 and 3), name of BCG (col. 4), large scale morphology of BCG (col.5), parsec-scale structure (col. 6) and references (col. 7). In particular, in column 3 we report the values of central mass accretion rate of the cluster derived from the literature (see col.7) in order to give an idea of the strength of the central gas density.
\begin{table*}
\caption{{\bf Results for the extended sample}: in the first column, we report the Abell cluster of BCG of our sample, in column 2, we indicate Y if the Abell cluster shows a cool core, N if it doesn't, in column 3 there are the value of central mass accretion rate of the cluster taken from the literature, in column 4 there are the names of BCGs. Column 5 is for the large scale morphology of the BCG: we use WAT for Wide Angle Tail radiosource, HT for Head Tail radiosource, MSO for medium symmetric source. In column 6, we mark the parsec-scale structure: one sided, two sided, core (unresolved) or n.d. for the non detections. In the last column, we give the references for the mas scale structure when the source is yet well studied at mas scale in literature and for values of Mass accretion rate given in column 3. }
\begin{center}
\small
\footnotesize
\label{tab:id}
\tabcolsep2mm
\begin{tabular}{ccccccc}
\hline
\hline
Abell Cluster&cool core&M$_{accr}$&BCG&Large scale&VLBI&Ref.\\
             &         & M$_{\odot}$/yr&         &    &    \\
\hline 
A400  &N  &0.0$^{+28.3}_{-0.0}$&3C75A          &WAT          &one sided&\citet{whi97}\\
      &N  &0.0$^{+28.3}_{-0.0}$&3C75B          &WAT          &one sided&\citet{whi97}\\
A407  &N  &4.6$^{+11.8}_{-4.6}$&UGC2489        &Tail rs      &one sided&\citet{whi97}\\
A539  &N  &2.1$^{+6.8}_{-2.1}$&UGC3274        &radio quiet  & n.d.&\citet{whi97}\\
A569  &N  &$>$5.2$^{+0.0}_{->4.2}$&NGC2329        &WAT          &one sided&\citet{whi97}\\
A576  &N  &17$^{+47}_{-17}$&CGCG261-059    &Tail rs      &one sided&\citet{whi97}\\
A690  &N  &0.0$^{+15.3}_{-0.0}$&B2 0836+29 II  &WAT          &one sided&\citet{gio05,whi97}\\
A779  &N  &3.1$^{+1.1}_{-1.1}$&NGC2832        &radio quiet  &n.d.&\citet{whi97}\\
A1185 &N  &0.0$^{+1.5}_{-0.0}$&NGC3550        &radio quiet  &n.d.&\citet{whi97}\\
A1213 &N  &0.0$^{+11.5}_{-0.0}$&4C29.41        & FRI         &one sided&\citet{whi97}\\
A1228 &N  &----&IC2738         &radio quiet  &n.d.&-\\
A1314 &N  &0.0$^{+3.0}_{-0.0}$&IC708          &WAT&one sided&\citet{whi97}\\
      &N  &0.0$^{+3.0}_{-0.0}$&IC712          &small WAT    &n.d.&\citet{whi97}\\
A1367 &N  &2.3$^{+6.8}_{-2.3}$&NGC3842        &small WAT    &n.d.&\citet{whi97}\\
      &N  &2.3$^{+6.8}_{-2.3}$&3C264          &HT           &one sided&\citet{la99,whi97}\\
A1656 &N  &0.0$^{+1.0}_{-0.0}$&NGC4874        &small WAT    &one sided&\citet{whi97}\\
A2147 &N  &0.0$^{+14.5}_{-0.0}$&UGC10143       &small WAT    & n.d.&\citet{whi97}\\
A2151 &N  &6.3$^{+26.3}_{-3.2}$&NGC6041        &small WAT    &core&\citet{whi97}\\
      &N  &6.3$^{+26.3}_{-3.2}$&NGC6047        &compact core+symmetric jets&n.d.&\citet{whi97}\\
A2162 &N  &----&NGC6086        &FRI, relic source &n.d&-\\
A2197 &N  &2.4$^{+3.0}_{-2.4}$&NGC6173        &point source &one sided&\citet{whi97}\\
A2634 &N  &0.0$^{+1.5}_{-0.0}$&3C465          &WAT          &one sided&\citet{ve95,whi97}\\
A2666 &N  &0.0$^{+2.6}_{-0.0}$&NGC7768        &Tail rs      &one sided&\citet{whi97}\\
A3526 &N  &5.2$^{+0.3}_{-0.3}$&PKS 1246-410      &small tailed rs&one-sided&\citet{tas06, hud09}\\
\\
A262 & Y &9.4$^{+21.2}_{-4.4}$&NGC708          &double-no core,jets &core&\citet{whi97}\\
A347 &SCF &7.8$^{+3.5}_{-2.7}$&NGC910          &radio quiet &n.d.&\citet{whi97}\\
A426 &Y   &291$^{+>-7}_{-58}$&3C84            &Compact core+Halo& two sided&\citet{tay06,whi97}\\
A780 &Y   &222$^{+98}_{-132}$&Hydra A         &double     &two sided&\citet{tay96,whi97}\\
A1795&Y   &321$^{+166}_{-213}$&4C26.42         &double     &two sided&\citet{liu09,whi97}\\
A2052&Y   &94$^{+84}_{-37}$&3C317           &bright core+halo (FRI)&two sided&\citet{ve04,whi97}\\
A2152& Y&20$^{+13}_{-20}$&UGC10187        &Tail rs&n.d.&\citet{whi97}\\
A2199&Y   &97$^{+9}_{-31}$&3C338           &double restarted&two sided&\citet{whi97,fe93}\\
A2390&Y   &247$^{+43}_{-91}$&B2151+174       &MSO&two sided&\citet{au06, all01}\\
A2597&Y   &501$^{+58}_{-512}$ &PKS 2322-123   &asymmetric radiosource (FRI)& two sided&\citet{tay99,che07}\\

\hline
\hline
\end{tabular}
\end{center}
\end{table*}

The extended sample is composed of 34 BCGs: 10 are in cool core
clusters and 24 are in non cool core clusters. Tab.5 summarises
statistically properties of the complete and the extended sample.

\begin{table*}[htp]
\caption{{\bf BCG counts in the complete (nearby) sample and expanded one}.  We
  report the number of BCG according to the cluster morphology and parsec-scale
  morphology. Note that most of the undetected sources in VLBA observations are 
in BCG that
  are radio quiet (or faint) in VLA observations (see \S 5.3).}
\begin{center}
\label{tab:id}
\tabcolsep2mm
\begin{tabular}{ccccccc}
\hline
\hline
Sample & Cluster  & Number & two-sided & one-sided & point & N.D. \\
       & morphology &        &           &           &       &      \\
\hline 
Complete & cool core    &  5     &  2 (40$\%$)       &   --      & 1     & 2    \\
         & non cool core    &  22    & --        &  12 (55$\%$)      & 1     & 9    \\
\hline
Expanded & cool core   &  10    & 7 (70$\%$)   & -- & 1  & 2  \\
         & non cool core  &  24    & --  & 14 (58$\%$) & 1  & 9  \\
\hline
\hline
\end{tabular}
\end{center}
\end{table*}

In the expanded sample, we find in cool core clusters:
\begin{center}
 \begin{itemize}
  \item 70$\%$ two-sided sources;
\item  20$\%$  non detected sources;
\item 10 $\%$ unresolved structures.
 \end{itemize}
\end{center}

Instead, in non cool core clusters, BCGs show:
\begin{center}
 \begin{itemize}
  \item 58$\%$ one-sided morphology;
\item 38$\%$ non detected sources;
\item 4$\%$ unresolved structures.
 \end{itemize}
\end{center}

\section{Discussion.}
The presence of a clear dichotomy between relaxed and non relaxed
clusters is evident (Table 5). At mas scale, one-sided structures are
only in BCGs in non cool core clusters, instead two-sided morphologies
are only in BCGs in cool core clusters.

\subsection{Jet velocities.}
We use as a comparison sample the Bologna Complete Sample (BCS)
\citep{gio01, gio05}. This sample is composed of 95 FRI radiogalaxies
spanning the same radio power range as our BCG sample. Moreover, it is
free of selection effects, in particular on jet velocity and
orientation.  As for our sample, VLBI observations at 5 GHz and
kiloparsec morphology information for most of these objects are
available and presented in \citet{liu09b}. Among the results from the
BCS study, \citet{gio05,liu09b} found that the one-sided jet
morphology is the predominant structure and only 22$\%$ of FRI radio
galaxies have two-sided jets. This is in agreement with expectations
based on a random orientation for sources with relativistic jets.
Based on conclusions for the BCS sample, we suggest that all FRIs 
outside of cool cores have
similar parsec-scale properties regardless of their host galaxy
classification (BCG or non BCG).  One-sided structures in non cool
core clusters are due to Doppler boosting effects in relativistic,
intrinsically symmetric jets.

Two-sided structures can be due either to relativistic jets in the
plane of the sky or to mildly relativistic jets. For our BCGs, we
exclude the first hypothesis as a consequence of statistical
considerations in comparison with the BCS results. It is not possible
that all BCGs in cool core clusters discussed here with resolved jets
are oriented in the plane of sky. Therefore we conclude that BCGs in
cool core clusters have on the parsec-scale mildly relativistic jets.
All resolved BCGs in cool core clusters show two-sided jets. This
result implies that BCGs in cool core clusters must be due to mildly
relativistic jets and they are not a consequence of relativistic jets
in the plane of sky.  To further test this hypothesis, in Fig.~10 we plot the observed total 
arcsecond radio 
power at
  408 MHz versus the observed arcsecond core radio power at 5 GHz
for all the sources of our extended sample. 
The
  solid black line is the correlation found by
  \citet{gio01} for sources with relativistic jets. 
According to the table 3, for UGC10143, UGC10187 and
  NGC6086 we draw the upper limits of the observed arcsecond core
  radio power at 5 GHz.
Despite the low number of sources discussed here 
we note that two-sided sources (crosses in the plot) detected 
in cooling clusters, and expected
to show mildly relativistic jets, are not in agreement with the
correlation found by \citet{gio01}, while
BCG with one-sided jets (dots in the figure) are in good agreement with the general correlation 
  confirming the presence of relativistic jets in these BCGs.

\subsection{Mildly relativistic jets.}

In BCGs at the center of cooling cores the gas density in the ISM
region is expected to be higher \citep{sal03}.  Studies of X-ray
emission of the hot intra-cluster medium (ICM) have pointed out the high
density of this gas in the central regions of many clusters.  For
example, estimated cooling rates of the order of 10 $M_\odot$/yr and
up to 100 $M_\odot$/yr implied that enormous quantities of material
should have accumulated (10$^{10}$ to 10$^{11}$ $M_\odot$ in a
fraction of a Hubble time).  Because of the dense ISM of BCGs in cool
core clusters we suggest that in BCGs in cool core clusters the jet
interaction with the ISM is already relevant on the parsec-scale.
 We note (see Table 4) that two-sided jets are present only in BCGs at the center of clusters with a central mass accretion rate $>$ 90  M$_{\odot}$/yr.

\citet{ro08} discussed the interaction between relativistic jets and
the surrounding ISM. They showed that a jet perturbation grows because
of Kelvin-Helmotz instability and produces a strong interaction of the
jet with the external medium with a consequent mixing and
deceleration.  The deceleration becomes more efficient as the density
ratio between the ambient medium and the jet increases. Light,
relativistic jets are expected in FR I sources, so the above effect
can slow them down from the parsec to the sub-kiloparsec-scale as
found in many sources \citep{tay96, ro08}. Since light jet beams imply
reduced jet kinetic powers, the model of \citet{ro08} leaves the
density contrast as the most likely candidate to account for the
discrepancies in the efficiency of the deceleration process. In this
scenario, as for sources in non cool core clusters, the jet begin
relativistic (and thus appears one-sided at the base) but a large
value of the density ratio can produce a sub-relativistic (and therefore
two-sided) heavy jet at a much shorter distance from the central 
engine as compared to ``normal'' FR I radio galaxies.

\subsubsection{Hydra A and 3C 84.}

There are two cases in particular, Hydra A and 3C\,84 in cool core
clusters A780 and A426 respectively (see \S4), where the above
scenario seems to be most evident.  Hydra A appears surprisingly
symmetric given the observed RM and depolarisation asymmetries seen on
large scales. \citet{tay96} suggested that the emission from the
symmetric parsec-scale jets is more dependent on interactions with the
surrounding material than on Doppler boosting.

3C\,84 is, in our sample, the nearest (z=0.0178) and best studied BCG at
high resolution. VLBA images at 86 GHz \citep{lee08} and 43 GHz
\citep{lis01} reveal that at a resolution of 0.32 mas ($\sim$ 0.07 pc)
the source is one-sided.  On larger scales, in VLBA data
\citep{tay06} 3C 84 shows two-sided morphology implying that the jet
interaction with the dense surrounding medium produces the slowing down
of the initially relativistic jet at sub-pc scale.

\section{Conclusions.}

BCGs are a unique class of objects. To study their properties on the
parsec-scale, we defined a complete sample selecting all BCGs in
nearby Abell clusters (DC$<$3) and declination $>$0$^{\circ}$. We
obtained VLBA observations at 5 GHz for these objects without radio
data at mas resolution available from literature. We find a different
behavior between BCGs in cool core and non cool core
clusters. Undetected and point like sources are found in BCGs of both
types of clusters. Undetected sources are generally a consequence of
low radio activity at the epoch of the observations (radio quiet core), 
or no radio emission whatsoever (radio quiet source).  
Point source morphologies may indicate insufficient sensitivity of
our data and/or core dominance effects. 

To better understand their properties, we added to our complete sample
other BCGs with detailed information about the radio emission at
parsec-scale and X-ray properties of the cluster. The extended sample
is composed of 34 BCGs: 11 in cool core clusters and 23 in non cool
core clusters. 
A dichotomy is found between the parsec-scale structures of BCGs in
cool core and non cool core clusters: all the resolved objects
($56\%$) in non cool core clusters show a one-sided jet, instead in
cool core clusters, all the resolved BCGs (64$\%$) show a two-sided
morphology.  Using the BCS sample as a comparison sample, we suggest
that one sided structure in non cool core clusters is due to Doppler
boosting effects in relativistic, intrinsically symmetric
jets. Furthermore, the dominance of two-sided jet structures only in
cooling clusters suggests sub-relativistic jet velocities. The
different jet properties can be related to a different jet origin or
to the interaction with a different ISM. In BCGs at the center of a
cooling core cluster the gas density in the ISM region is expected to
be higher. Therefore we can assume a strong interaction of the jet at
parsec resolution with the environment. However, a large value of the
density ratio of the medium to the jet, can produce entrainment 
leading to a sub-relativistic
and heavy jet at a much shorter distance (pc scale). This suggestion
is supported by data from the literature on Hydra A and 3C 84, BCGs of two cool
core clusters (A780 and A426 respectively).  We also found episodic
jet activity from the central engine of AGN in a few objects. The
recurrent activity of the radio source in cool core clusters is of great
interest to the study of AGN feedback in clusters.

More data are necessary to better understand and test the nature of
the difference that we note between BCGs in cool and non cool
clusters. We would also like to understand the properties of the
restarted emission in BCGs.  To improve the statistic, observations of
a larger sample of BCGs in cooling and relaxed clusters with the VLBA
is necessary.

\begin{acknowledgements}
We thank the staff of NRAO involved in the observations for their
assistance. NRAO is a facility of the National Science Foundation, operated
under cooperative agreement by Associated Universities, Inc. This
research has made use of the NASA/IPAC Extragalactic Data Base (NED),
which is operated by the JPL, California Institute of Technology,
under contract with the National Aeronautics and Space Administration. {\bf We thanks also an anoymous Referee for useful comments which improved this work.}
\end{acknowledgements}
\vfill\eject

\vfill\eject

\onecolumn

\begin{figure}
\centering
\includegraphics[width=0.45\textwidth]{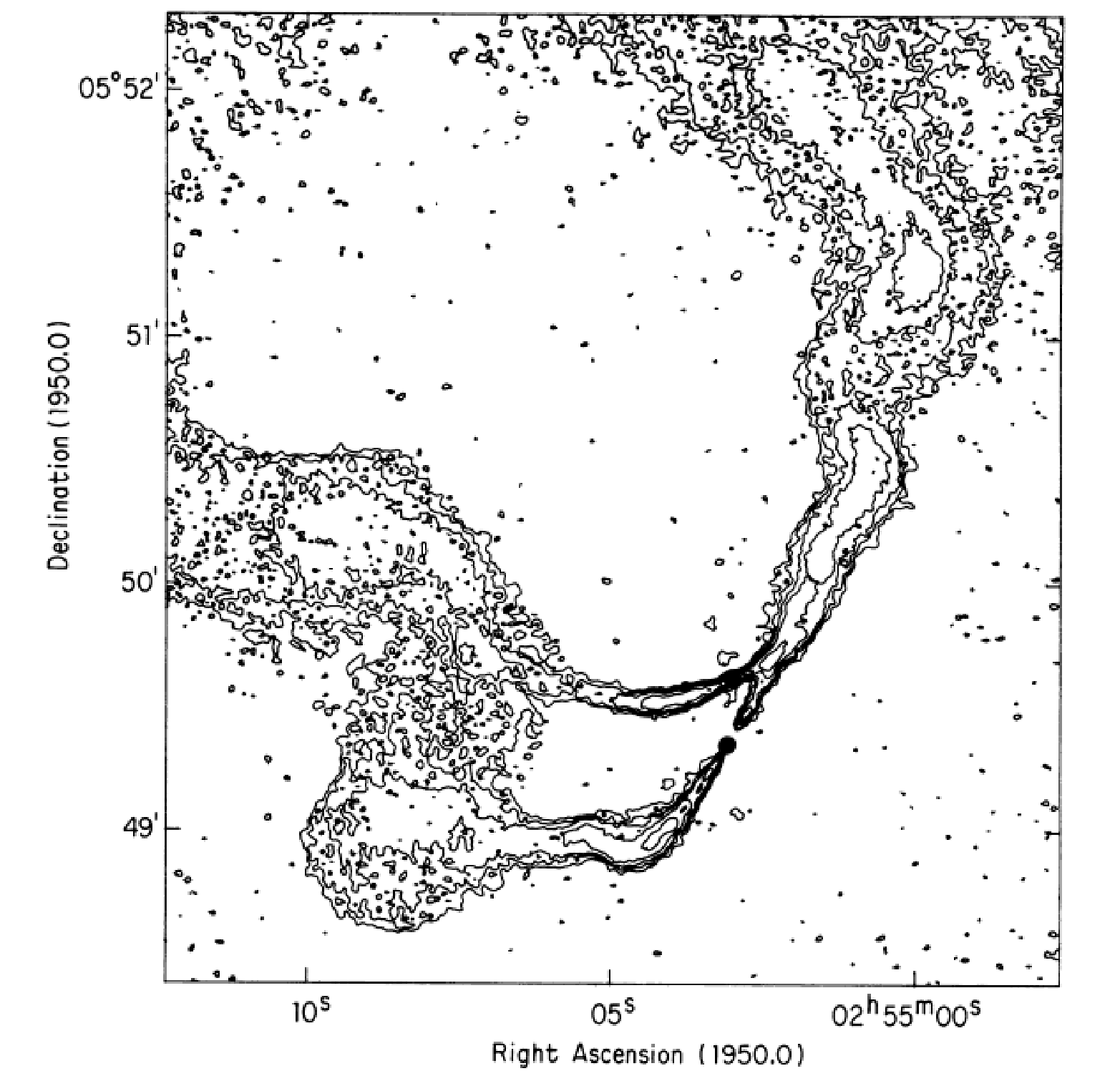}
\hfill
\includegraphics[width=0.30\textwidth, angle=-90]{el2.ps}
\hfill
\includegraphics[width=0.30\textwidth, angle=-90]{el3.ps}
\caption{The double BCG in the non cool core cluster A400. Left: VLA
  image at 5 GHz of 3C\,75A and 3C\,75B from \citet{ow85}. Contour
  intervals are (-1, 1, 2, 4, 8. 16, 32, 64, 128, 256, 512) $\times$
  0.1 mJy per clean beam. Center: one-sided 5 GHz VLBA tappered image
  of 3C\,75A. Contour levels are -0.3, 0.3, 0.6, 1.2, 2.4, 4.8, 9.6 and
  19.2 mJy/beam. The peak flux density is 19.9 mJy/beam, the noise
  level is 0.1 mJy/beam and the restoring beam is 2 $\times$ 2
  mas$^{2}$, P.A. = 0 $^{\circ}$. Right: one-sided 5 GHz VLBA tappered
  image of 3C\,75B. Contour levels are -0.3, 0.3, 0.6, 1.2, 2.4, 4.8,
  9.6 and 19.2 mJy/beam. The peak flux density is 37.2 mJy/beam, the
  noise level is 0.1 mJy/beam and the restoring beam is 4 $\times$ 4
  mas$^{2}$, P.A. = 0 $^{\circ}$. }
\end{figure}

\begin{figure}
\centering
\includegraphics[width=0.3\textwidth]{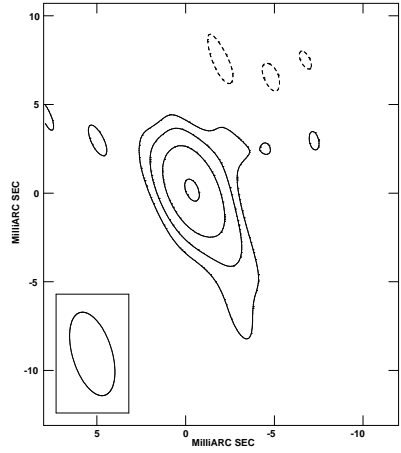}
\caption{The 5 GHz VLBA naturally weighted image of UGC2489, BCG of
  A407. Contour levels are -0.31, 0.31, 0.62, 1.24 and 2.48
  mJy/beam. The peak flux density is 2.6 mJy/beam, the noise level is
  0.1 mJy/beam and the restoring beam is 4.87 $\times$ 2.23 mas$^{2}$,
  P.A. = 16.8 $^{\circ}$.  }.\label{fig1}
\end{figure}

\begin{figure}
\centering
\includegraphics[width=0.3\textwidth, angle =-90]{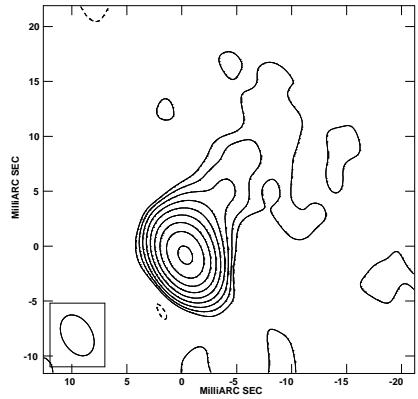}
\caption{5 GHz VLBA tappered map of NGC2329, BCG of A569. Contour
  levels are -0.23 0.23 0.46 0.92 1.84 3.68 7.36 14.72 29.44 and 58.88
  mJy/beam. The peak flux density is 66.8 mJy/beam, the noise level is
  0.08 mJy/beam and the restoring beam is 4.0 $\times$ 2.7 mas$^{2}$,
  P.A. = 30 $^{\circ}$.  }.\label{fig1}
\end{figure}

\begin{figure}
\centering
\includegraphics[width=0.3\textwidth]{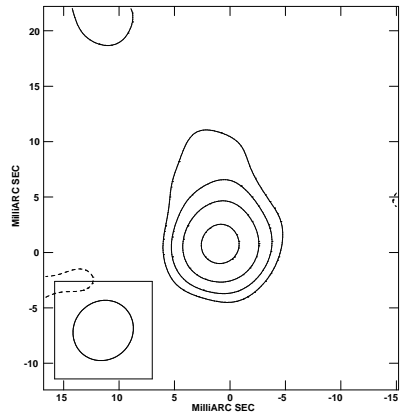}
\caption{The 5 GHz VLBA tappered map of CGCG261-059, BCG of A576. Contour
  levels are -0.3, 0.3, 0.6, 1.2 and 2.4 mJy/beam. The peak flux
  density is 3.0 mJy/beam, the noise level is 0.1 mJy/beam and the
  restoring beam is 5.63 $\times$ 5.24 mas$^{2}$, P.A. = - 45.19
  $^{\circ}$.  }.
\end{figure}

\begin{figure}
\centering
\includegraphics[width=0.3\textwidth, angle =-90]{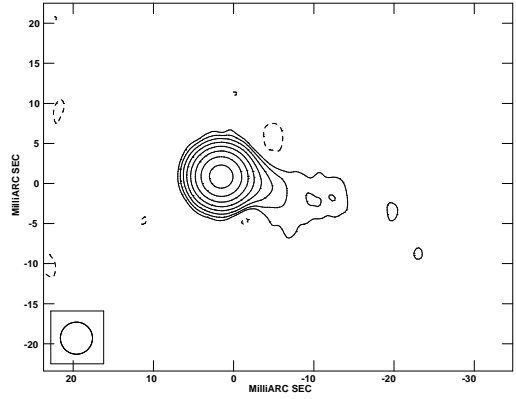}
\caption{The 5 GHz VLBA tappered map of 4C29.41, BCG of A1213. Contour
  levels are -0.2, 0.2, 0.4, 0.8, 1.6, 3.2, 6.4, 12.8 and
  25.6. mJy/beam. The peak flux density is 36.7 mJy/beam, the noise
  level is 0.07 mJy/beam and the restoring beam is 4 $\times$ 4
  mas$^{2}$, P.A. = 0 $^{\circ}$. }
\end{figure}

\begin{figure}
\centering
\includegraphics[width=0.3\textwidth]{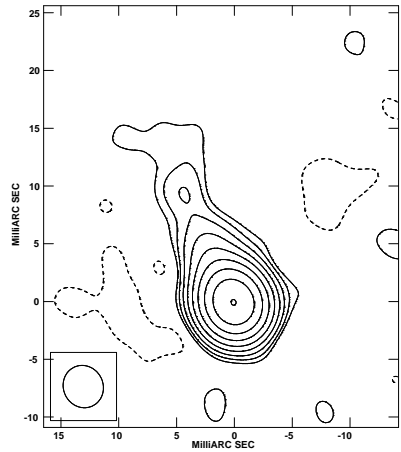}
\caption{The 5 GHz VLBA tappered map of IC708, BCG of the second
  condensation of the binary cluster A1314. Contour levels are -0.15,
  0.15, 0.3, 0.6, 1.2, 2.4, 4.8, 9.6, 19.2 and 38.4 mJy/beam . The
  peak flux density is 38.8 mJy/beam, the noise level is 0.05 mJy/beam
  and the restoring beam is 3.71 $\times$ 3.43 mas$^{2}$, P.A. = 25.4
  $^{\circ}$.  }
\end{figure}

\begin{figure}
\centering
\includegraphics[width=0.3\textwidth, angle =-90]{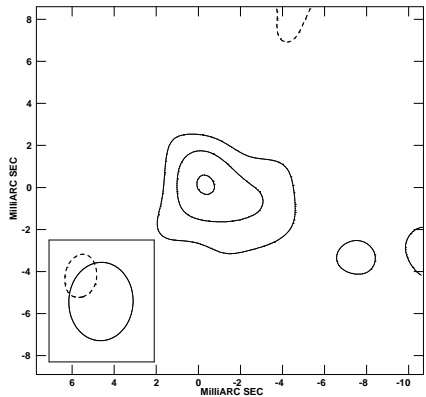}
\caption{The 5 GHz VLBA tappered map of NGC4874, BCG of A1656. Contour
  levels are -0.24 0.24 0.48 and 0.76 mJy/beam. The peak flux density
  is 0.8 mJy/beam, the noise level is 0.08 mJy/beam and the restoring
  beam is 3.73 $\times$ 3.05 mas$^{2}$, P.A. = -2.1 $^{\circ}$.  }
\end{figure}

\begin{figure}
\centering
\includegraphics[width=0.3\textwidth, angle =-90]{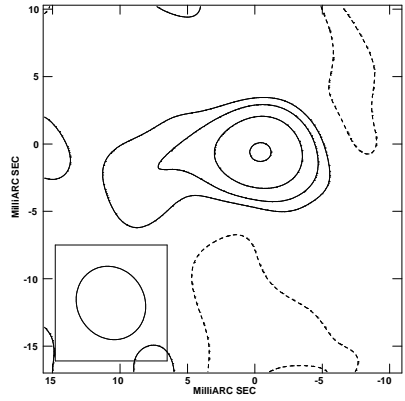}
\caption{The 5 GHz VLBA tappered map of NGC6173, BCG of A2197. Contour
  levels are -0.19 0.19 0.38 0.76 and 1.52 mJy/beam. The peak flux
  density is 1.6 mJy/beam, the noise level is 0.06 mJy/beam and the
  restoring beam is 5.58 $\times$ 5.0 mas$^{2}$, P.A. = 30 $^{\circ}$.
}
\end{figure}

\begin{figure}
\centering
\includegraphics[width=0.3\textwidth]{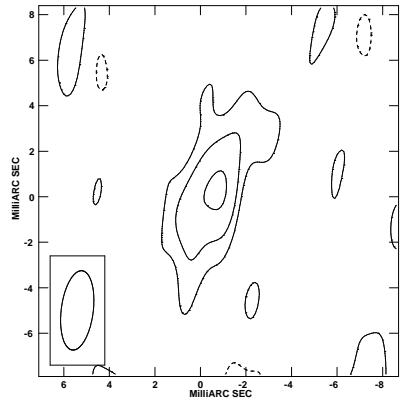}
\caption{The 5 GHz VLBA naturally weighting map of NGC7768, BCG of
  A2666. Contour levels are -0.17 0.17 0.34 and 0.64 mJy/beam. The
  peak flux density is 0.72 mJy/beam, the noise level is 0.09 mJy/beam
  and the restoring beam is 3.51 $\times$ 1.40 mas$^{2}$, P.A. = -7.15
  $^{\circ}$. }
\end{figure}

\begin{figure}
\centering
\includegraphics[width=0.9\textwidth]{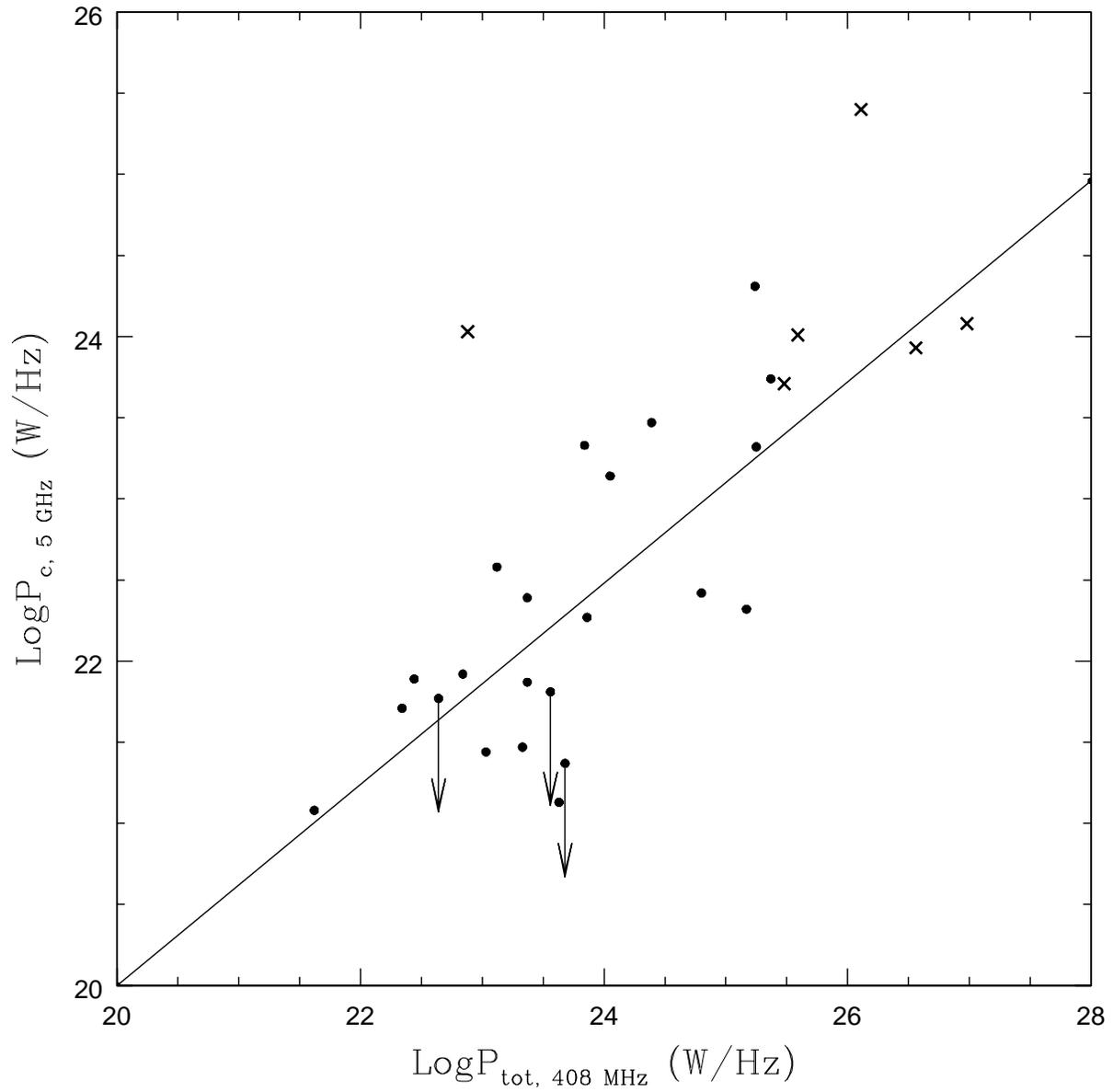}
\caption{Total arcsecond radio power at 408 MHz versus observed
  arcsecond core radio power at 5 GHz for all BCGs of our complete
  sample with available information. The solid line is the relation
  found by \citet{gio01}. Crosses (x) represent BCGs with two-sided
pc scale jets and dots, one-sided sources.}

\end{figure}

\end{document}